\documentclass[groupedaddress,aps,pra,superscriptaddress,showpacs,twocolumn,prl]{revtex4}%
\usepackage{epsfig,dsfont,amssymb,amsmath,amsthm,amsfonts,amsbsy,mathrsfs}
\usepackage{graphicx, color}
\usepackage{epstopdf}
\usepackage{mathdots}
\usepackage{float}
\begin{document}

\title{Superactivation of monogamy relations for nonadditive quantum correlation measures}

\author{Zhi-Xiang Jin}
\thanks{Corresponding author: jzxjinzhixiang@126.com}
\affiliation{School of Mathematical Sciences,  Capital Normal University,  Beijing 100048,  China}
\author{Shao-Ming Fei}
\thanks{Corresponding author: feishm@cnu.edu.cn}
\affiliation{School of Mathematical Sciences,  Capital Normal University,  Beijing 100048,  China}
\affiliation{Max-Planck-Institute for Mathematics in the Sciences, Leipzig 04103, Germany}

\begin{abstract}
We investigate the general monogamy and polygamy relations satisfied by quantum correlation measures.
We show that there exist two real numbers $\alpha$ and $\beta$ such that for any quantum correlation measure $Q$, $Q^x$ is monogamous if $x\geq \alpha$ and polygamous if $0\leq x\leq \beta$ for a given multipartite state $\rho$. For $\beta <x<\alpha$, we show that the monogamy relation can be superactivated by finite $m$ copies $\rho^{\otimes m}$ of $\rho$ for nonadditive correlation measures.
As a detailed example, we use the negativity as the quantum correlation measure to illustrate such superactivation of monogamy properties. A tighter monogamy relation is presented at last.
\end{abstract}

\maketitle

\section{INTRODUCTION}

Quantum mechanically the more a two-level system is entangled with another two-level system, the less it can be entangled with a third one \cite{VJW}. This behavior, known as entanglement monogamy, has also been found in larger systems \cite{FMA,KSS,HPB,HPBB,JIV,CYS}. Monogamy of entanglement is one of the nonintuitive phenomena of quantum physics that distinguish it from classical physics. Classically, three random bits can be maximally correlated. However, it is not possible to prepare three qubits in a way that any two qubits are maximally entangled \cite{VJW}, i.e., a quantum system entangled with one of other subsystems limits its entanglement with the remaining ones. Moreover, the monogamy property has emerged as the ingredient in the security analysis of quantum key distribution \cite{MP}.

The monogamy of entanglement has fundamental implications in some quantum information processing. For example, the lack of monogamy is considered a huge obstacle to the implementation of quantum cryptography \cite{jmr,ml}. It also plays roles in detecting phases in many-body physics \cite{max,kjm}, and provides information that may help us to understand the mysterious behavior of black holes \cite{aad}. Moreover, the monogamy of quantum correlation is essential for proving that asymptotic cloning is equivalent to state estimation \cite{jbe} and making quantum key distribution secure \cite{vss}.

For a tripartite system $A$, $B$ and $C$, the usual monogamy of a quantum correlation measure $Q$ implies that the correlation $Q_{A|BC}$ between $A$ and $BC$ satisfies the following inequality,
\begin{eqnarray}\label{monogamy}
Q_{A|BC}\geq Q_{AB} +Q_{AC},
\end{eqnarray}
where $Q_{AB}$ ($Q_{AC}$) stands for the correlation between systems $A$ and $B$ ($C$) of the corresponding reduced bipartite system.
Inequality (\ref{monogamy}) was originally proven for arbitrary three-qubit states, adopting the squared concurrence $C^2$ as the correlation (entanglement) measure \cite{VJW}. Variations of the CKW inequality and its generalizations to $N$-partite systems have been presented for a number of entanglement measures in discrete as well as continuous variable systems \cite{byk1,ZXN,byk2,JZX,jll,j012334,042332,gy1}.

Dually, the polygamy relation is quantitatively described by
\begin{eqnarray}\label{polygamy}
Q_{A|BC}\leq Q_{AB} +Q_{AC}.
\end{eqnarray}
Recently, polygamy inequalities have been given for multiqubit systems under various entanglement measures \cite{fgj,sa,jzx1,jzx2,gy2}. The monogamy and polygamy inequalities are also proposed in terms of
non-negative power of various entanglement measures \cite{JZX,jll,j012334}, and a generalization of the polygamy constraint of multipartite entanglement in arbitrary dimensional quantum systems has been given in Ref. \cite{042332}.

It has been shown that the concurrence, negativity, and tangle adopt monogamy relations for multiqubit systems \cite{ZXN,fh,bcs,ycs}. However, it is still unclear for high dimensional systems. It has been shown that the entanglement of assistance and the entanglement of assistance associated with the Tsallis-$q$ entropy are polygamous for any multipartite systems \cite{fgj,062302,gy1,gpw}. The concurrence of assistance, tangle of assistance and negativity of assistance are proved to be polygamous only for multiqubit systems \cite{GSB,bcs,fsm}. The polygamy problem for other entanglement measures remains open for high dimensional systems.

Due the importance of monogamy relations, it is also interesting to ask whether the monogamy property of a quantum state can be superactivated. Namely, if $\rho$ does not satisfy monogamous relations with respect to some correlation measure, can the many copy state $\rho\otimes\rho\cdots\otimes\rho$ be monogamous?

In this paper, we study the monogamy and polygamy relations for general quantum correlation measures. We show that there always exist two real numbers $\alpha$ and $\beta$ for any quantum correlation measures and
any given quantum states: $Q^x$ is monogamous if $x\geq \alpha$ and polygamous if $0\leq x\leq \beta$.
For $\beta <x<\alpha$, we find that the monogamy property depends not only on the quantum state, the quantum correlation measure, the power of the quantum corrlation measure, but also on the number of copies of the state. The phenomena is similar to the nonlocality revealed by the violation of Bell inequalities.
For some bipartite states that admit local hidden variable models, their nonlocality can be super activated by $m$ copies of the states \cite{supera}. Here we show that for a state which does not admit monogamous relations, the $m$ copies of the state could satisfy monogamous relations. We use the negativity as the quantum correlation measure to illustrate such super activation of monogamy properties.
We also present the definition of regularized quantum correlation measures and show that almost all regularized quantum correlation measures satisfy the monogamy relations, except for the measures which are additive. We give a tighter monogamy relation at last and generalize it to the multipartite case.

Throughout this paper, $Q(\rho_{A|B_1B_2,\cdots,B_{N-1}})$ denotes the quantum correlation of the state $\rho_{AB_1B_2,\cdots,B_{N-1}}$ under bipartite partition  $A$ and $B_1B_2,\cdots,B_{N-1}$, which keeps invariant under discarding subsystems only for states satisfying  monogamy relations. For simplicity, we denote $Q(\rho_{AB_i})$ by ${Q}_{AB_i}$, and ${Q}(\rho_{A|B_1B_2,\cdots,B_{N-1}})$ by ${Q}_{A|B_1B_2,\cdots,B_{N-1}}$.

\section{Monogamy and polygamy relations for quantum correlation measures}

{\bf[Theorem 1]}. Let $Q$ be a continuous measure of quantum correlation. For any tripartite state $\rho_{ABC}$ in discrete finite dimensional Hilbert space, there exist real numbers $\alpha$
and $\beta$ such that

(1) if $x\geq \alpha$, then $Q^x$ is monogamous,
\begin{eqnarray}\label{th11}
Q^x_{A|BC}\geq Q^x_{AB} +Q^x_{AC},
\end{eqnarray}

(2) if $0\leq y\leq \beta$, then $Q^y$  is polygamous,
\begin{eqnarray}\label{th12}
Q^y_{A|BC}\leq Q^y_{AB} +Q^y_{AC}.
\end{eqnarray}

{\sf[Proof]}.
As $Q$ is a measure of quantum correlation, it is nonincreasing under partial trace. Therefore, $Q_{A|BC}\geq \max\{Q_{AB},Q_{AC}\}$ for any state $\rho_{ABC}$. If $Q_{A|BC}=0$, the result is obvious. Therefore, we assume $Q_{A|BC}> 0$ and set $x_1=Q_{AB}/Q_{A|BC}$, $x_2=Q_{AC}/Q_{A|BC}$. Clearly, there exists $\gamma$ such that
\begin{eqnarray}\label{pfth11}
x_1^\gamma+x_2^\gamma\leq 1,
\end{eqnarray}
since $x_1$ and $x_2$ decrease when $\gamma$ increases. Let $f(\rho_{ABC})$ be the smallest value of $\gamma$ that saturates the inequality (\ref{pfth11}). As $Q$ is continuous, we obtain
$\alpha=\sup_{\rho_{ABC}}f(\rho_{ABC})$, which proves the case 1 of Theorem 1.

On the other hand, there always exists $\delta\geq0$ such that
\begin{eqnarray}\label{pfth13}
x_1^\delta+x_2^\delta\geq 1,
\end{eqnarray}
since $x_1$ and $x_2$ increase when $\delta$ decreases. Let $g(\rho_{ABC})$ be the largest value of $\delta$ that saturates the inequality (\ref{pfth13}). As $Q$ is continuous, we obtain
$\beta=\inf_{\rho_{ABC}}g(\rho_{ABC})>0$, which proves the case 2 of Theorem 1.
\hfill \rule{1ex}{1ex}

Note that the monogamy relation satisfied by $Q^\alpha$ remains for larger power than $\alpha$ \cite{ZXN}, i.e., $Q^\alpha_{A|BC}\geq Q^\alpha_{AB} +Q^\alpha_{AC}$ implies that $Q^\eta_{A|BC}\geq Q^\eta_{AB} +Q^\eta_{AC}$ for any $\eta\geq \alpha$. This power $\alpha$ depends on the correlation measure $Q$ and the dimension of the system \cite{oyc}. Generally, it is hard to compute, especially for higher dimensional systems. The case 1 of Theorem 1 indicates that almost any correlation measures can give rise to monogamous relations for sufficient larger $\alpha$. Similarly, the polygamy relation of $Q^\beta$ can be preserved for smaller power \cite{ak}, i.e., $Q^\beta_{A|BC}\leq Q^\beta_{AB} +Q^\beta_{AC}$ implies that $Q^\theta_{A|BC}\leq Q^\theta_{AB} +Q^\theta_{AC}$ for any $0\leq \theta\leq \beta$. The polygamy power $\beta$ also depends on the measure $Q$. The monogamy power $\alpha$ is a dual concept of the polygamy power $\beta$. They both reflect the shareability of the correlations $Q$ among the subsystems. In the following, we consider the monogamy relation of $Q^x$ for $\beta\leq x\leq \alpha$.

\section{Superactivation of monogamy relations}

Monogamy characterizes the distribution of quantum correlations, which depends on the quantum states, the quantum correlation measure $Q$, the dimension of the system, and the power $x$ in the quantum correlation measure $Q^x$. In Refs. \cite{ZXN,tj,jll,wj} the monogamy relations are established for $x\geq 2$ in multiqubit systems for concurrence and negativity, while for $x\geq\sqrt{2}$ for the entanglement of formation. However, it has been proved that the monogamy inequality fails for some three-qutrite quantum states under the squared concurrence.

From Theorem 1 it is obvious that there exist states $\rho$ such that for $\beta\leq x\leq \alpha$, $Q^x$ does not satisfy a monogamy relation. To investigate the monogamy property for such states,
we define, from the inequality (\ref{monogamy}), the residual quantum correlation as
$R(\rho_{ABC})=Q_{A|BC}- Q_{AB}-Q_{AC}$
for the tripartite case. Such residual correlation increases when $Q$ is replaced by $Q^x$ and $x$ increases \cite{jll,ksb}, a fact that can also be deduced from Theorem 1. In other words, the residual quantum correlation $R$ can be changed from negative to positive by varying $x$.
This fact enables the super activation of monogamy relations, similar to super activation of nonlocality \cite{supera}:
A state $\rho_{ABC}$ does not satisfy the monogamy relation, while the state $\rho^{\otimes m}_{ABC}$ is monogamous, i.e., $R(\rho^{\otimes m}_{ABC})\geq0$ while $R(\rho_{ABC})\leq0$ for some $m>1$.

Here, $\rho^{\otimes m}_{A_1A_2...A_n}=\rho_{A_{11}A_{12}...A_{1m}A_{21}...A_{2m}...A_{n1}...A_{nm}}$, $A_{ij}$ denotes the $i$th party of the $j$th copy of $\rho_{A_1A_2...A_n}$. From the definition of the residual quantum correlation, one gets $R^{(n)}(\rho^{\otimes m}_{A_1A_2...A_n})=Q(\rho^{\otimes m}_{A_1|A_2...A_n})-\sum_{i=2}^nQ(\rho^{\otimes m}_{A_1A_i})$, where $\rho^{\otimes m}_{A_1A_i}=\rho_{A_{11}...A_{1m}A_{i1}...A_{im}}=\mathrm{tr}_{A_{21}...A_{(i-1)m}A_{(i+1)1}...A_{nm}}(\rho_{A_{11}A_{12}...A_{1m}A_{21}...A_{2m}...A_{n1}...A_{nm}})$, $Q(\rho^{\otimes m}_{A_1|A_2...A_n})=Q(\rho_{A_{11}A_{12}...A_{1m}|A_{21}...A_{2m}...A_{n1}...A_{nm}})$ denotes the quantum correlation between the first party and the rest ones after the $m$ copies of $\rho_{A_1A_2...A_n}$, i.e., the quantum correlations between $A_1$ and $\bar{A_1}$, view $A_{11}A_{12}...A_{1m}$ and $A_{21}...A_{2m}...A_{n1}...A_{nm}$ as $A_1$ and $\bar{A_1}$, respectively, and $Q(\rho^{\otimes m}_{A_1A_i})=Q(\rho_{A_{11}...A_{1m}|A_{i1}...A_{im}})$ is the quantum correlation of the first party and the $i$th party.

In the following, we consider the quantum correlation measures that are nonadditive. If a measure of quantum correlation is additive, i.e., $Q(\rho\otimes \delta)=Q(\rho)+Q(\delta)$, then a nonmonogamous state cannot be changed to a monogamy one by superactivation. We have the following result on superactivation of monogamy relations.

{\bf[Theorem 2]}. Let $Q$ be a continuous measure of quantum correlation. For any state $\rho_{ABC}$ that $Q^x$ does not satisfy the monogamy relation (\ref{th11}) for $\beta\leq x\leq \alpha$, there always exists a positive integer $m$ such that
\begin{eqnarray}\label{th2n}
Q(\rho^{\otimes m}_{A|BC})\geq Q(\rho^{\otimes m}_{AB}) +Q(\rho^{\otimes m}_{AC}).
\end{eqnarray}

{\sf[Proof]}. For any state $\rho_{AB}$ of the composite systems $A$ and $B$ with eigenvalues $\lambda_i$
and the corresponding eigenstates $|i\rangle$, let us introduce a third system $B'$ such that $|\psi\rangle=\sum_i\sqrt{\lambda_i}|i\rangle|\hat{i}\rangle$ is a pure state of the tripartite system $ABB'$, with orthonormal basis $|\hat{i}\rangle$ of $B'$.
Then $\rho_{AB}=\mathrm{tr}_{B'}|\psi\rangle\langle\psi|$, where $\mathrm{tr}_{B'}$ is the partial trace over $B'$. As $\mathrm{tr}_{B'}$ is a local operation performed on $B'$, $Q(\rho_{AB})\leq Q(|\psi\rangle\langle\psi|)$. Since $\rho_A=\mathrm{tr}_{BB'}|\psi\rangle\langle\psi|=\mathrm{tr}_B\rho_{AB}$, the Schmidt coefficients of $|\psi\rangle$ are $\sqrt{\lambda_i(\rho_A)}$. Hence, the quantum correlation has the form, $Q(|\psi\rangle\langle\psi|)=f(\vec{\lambda}(\rho_A))$, where $f$ is a function of $\vec{\lambda}(\rho_A)$ given by the nonzero eigenvalues of the state $\rho_A$. Thus, $Q(\rho_{AB})$ depends on $\rho_A$ only and $Q(\rho_{AB})\leq f(\vec{\lambda}(\rho_A))$. Therefore, there exists a positive number $0\leq L\leq 1$ such that $Q(\rho_{AB})=L f(\vec{\lambda}(\rho_A))$, and $Q^k(\rho_{AB})=L^k f^k(\vec{\lambda}(\rho_A))=g(\vec{\lambda}^k(\rho_A))$, where $k$ is a positive integer, $\vec{\lambda}^k(\rho_A)=(\lambda_1^k(\rho_A),...,\lambda_{r_A}^k(\rho_A))$, and $g$ is a function of  $\lambda_i^k(\rho_A)$, $i=1,2,...,r_A$, $r_A$ is the rank of the reduced density matrix $\rho_A$.
Let us now consider $\rho^{\otimes m}_{AB}$. Since the eigenvalues of $\rho^{\otimes m}_{AB}$ are $\{\lambda_i^m\}$, $i=1,2,..., r_{AB}$, $r_{AB}$ is the rank of density matrix $\rho_{AB}$, which are the elements of the function $Q^m(\rho_{AB})$. Using the Theorem 1, we have that there always exits a positive integer $m$ such that the inequality (\ref{th2n}) holds.\hfill \rule{1ex}{1ex}

As an example let us consider the three qubits $W$ state, $|W\rangle_{ABC}=\frac{1}{\sqrt{3}}(|100\rangle+|010\rangle+|001\rangle)$, which does not satisfy the usual monogamy relation (\ref{monogamy}) \cite{JZX,jll}.
We consider the superactivation of monogamy relations for $W$ states under the entanglement measure negativity. Given a bipartite state $\rho_{AB}$, the negativity is defined by \cite{GRF}
$N(\rho_{AB})=(||\rho_{AB}^{T_A}||-1)/2$,
where $\rho_{AB}^{T_A}$ is the partial transpose with respect to the subsystem $A$, $||X||$ denotes the trace norm of $X$, i.e., $||X||=\mathrm{Tr}\sqrt{XX^\dag}$.
Negativity is a computable measure of entanglement, and is a convex function of $\rho_{AB}$. For any bipartite pure state $|\psi\rangle_{AB}$ in $d\otimes d'$ $(d\leq d')$ quantum system with Schmidt decomposition $|\psi\rangle_{AB}=\sum_{i=0}^{d-1}\sqrt{\lambda_i}|ii\rangle$, $\lambda_i\geq0$, $\sum_{i=0}^{d-1}\lambda_i=1$, the negativity $N(\rho_{AB})$ is given by
$N(|\psi\rangle_{AB})=2\sum_{i<j}\sqrt{\lambda_i\lambda_j}=(\mathrm{Tr}\sqrt{\rho_A})^2-1$,
where $\lambda_i$ are the eigenvalues of the reduced density matrix of $|\psi\rangle_{AB}$. For a mixed state $\rho_{AB}$, the convex-roof of entanglement negativity (CREN) is defined by
\begin{eqnarray}\label{nc}
N_c(\rho_{AB})=\mathrm{min}\sum_ip_iN(|\psi_i\rangle_{AB}),
\end{eqnarray}
where the minimum is taken over all possible pure state decompositions $\{p_i,~|\psi_i\rangle_{AB}\}$ of $\rho_{AB}$. From Refs. \cite{bcs,ly}, $N^{x}_c$ satisfies the monogamy relation if $x\geq2$ for states in $2\otimes 2\otimes 2^n$ systems.
It is easy to get that $x=2$ for the $W$-class states \cite{JZX,jll}.
$N_c$ (the case $x=1$) itself is not monogamous for the $W$ states.
Consider $m$ copies of the $W$ states.
We have $N_c(|W\rangle_{A|BC}^{\otimes m})=\frac{1}{2}[(1+\frac{4\sqrt{2}}{3})^m-1]$, $N_c(\rho_{AB}^{\otimes m})=N_c(\rho_{AC}^{\otimes m})=\frac{1}{2}[(1+\frac{4}{3})^m-1]$, where $\rho_{AB}$ and $\rho_{AC}$ are the reduced states of $|W\rangle_{ABC}$. The relation between
$N_c(|W\rangle_{A|BC}^{\otimes m})$ and the summation of $N_c(\rho_{AB}^{\otimes m})$ and $N_c(\rho_{AC}^{\otimes m})$ is shown in Fig. 1. From Fig. 1 one can see that $|W\rangle_{ABC}^{\otimes m}$ satisfy the monogamy inequality for $m>3$.

\begin{figure}
  \centering
  \includegraphics[width=9cm]{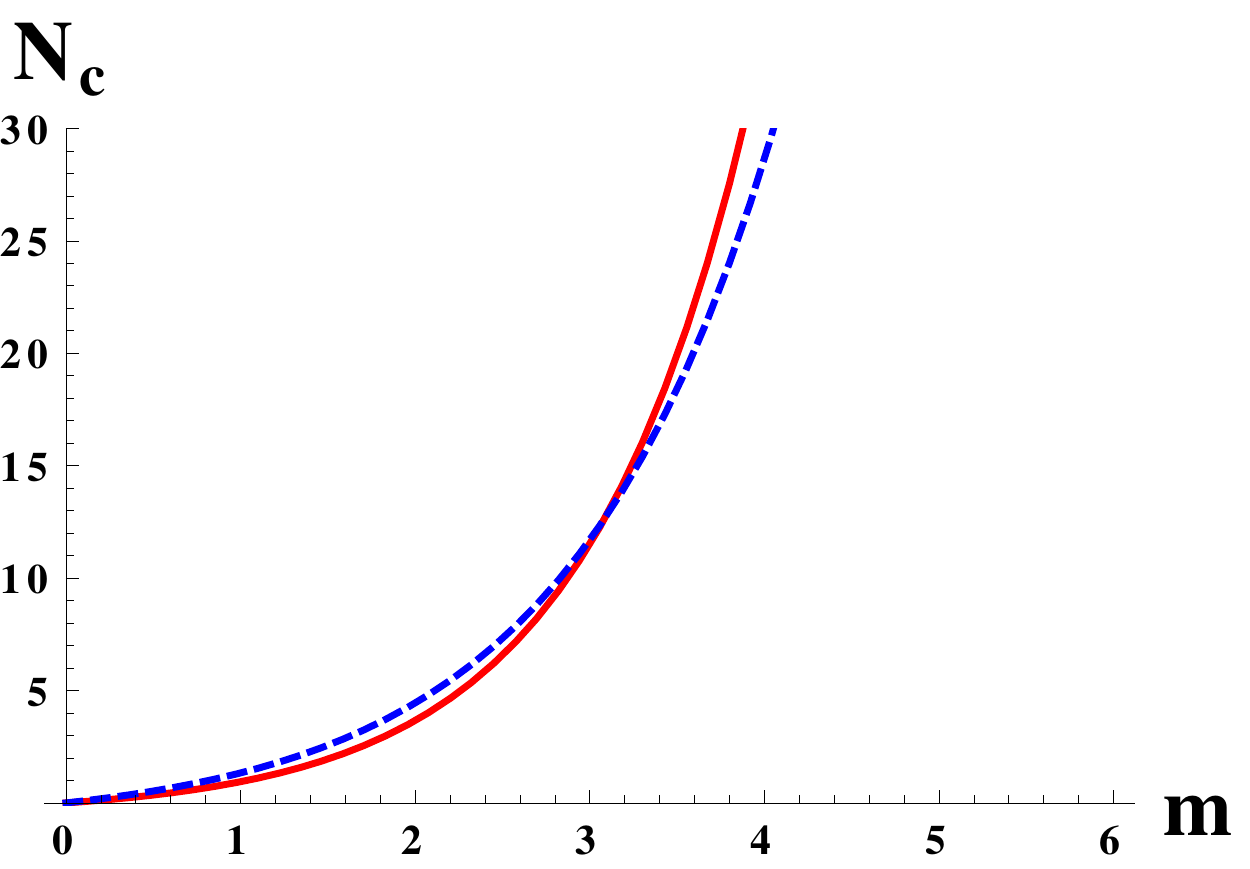}\\
  \caption{Solid (red) line is the negativity $N_c$ of $|W\rangle_{ABC}^{\otimes m}$; dashed (blue) line is the sum of the negativity $N_c$ of $\rho_{AB}^{\otimes m}$ and $\rho_{AC}^{\otimes m}$.}\label{1}
\end{figure}

Similar to violation of Bell inequalities, which can be enhanced by superactivation, the monogamy relations can also be superactivated. The monogamy relation satisfied by the $x$th power of a quantum correlation measure $Q$ can be revealed by finite copies of a state for $\beta<x<\alpha$. We can define the regularized quantum correlation measure of $Q$ as $Q^\infty(\rho)=\lim_{m\to\infty}\frac{1}{m}Q(\rho^{\otimes m})$. Thus, from the conclusion of Theorem 2, it is easy to get $Q^\infty(\rho_{A|BC})\geq Q^\infty(\rho_{AB})+Q^\infty(\rho_{AC})$.

It is clear that monogamy is a property of both the quantum state and quantum correlation measure. From the results of Theorem 1 and Theorem 2, the monogamy relation may also depend on the power of the correlation measure and the number of copies. However, not all quantum correlation measures are able to make a quantum state from being nonmonogamous to monogamous by finite copies, such as any quantum correlation measures satisfying the property of additivity, $Q(\rho^{\otimes m})=m\,Q(\rho)$. Moreover, there are special classes of states, which are always monogamous for any quantum correlation measures. For example, the generalized $n$-partite GHZ-class state admitting the multipartite Schmidt decomposition \cite{gdl,gjl}, $|\psi\rangle_{A_1A_2\cdots A_n}=\sum_{i=1}\lambda_i|i_1\rangle\otimes|i_2\rangle\otimes\cdots \otimes|i_n\rangle$, $\sum_i\lambda_i^2=1$, $\lambda_i>0$, for which we always have $E(|\psi\rangle_{A_1|A_2\cdots A_n})>0$, while $E(\rho_{A_1A_i})=0$ for all $i=2,\cdots,n$, for any quantum entanglement measures $E$.

{\sf Remark}. In Ref. \cite{supera}, the authors have shown that quantum nonlocality can be superactivated. For some bipartite states that admit local hidden variable models, their nonlocality can be superactivated by copies of the states. There are quantum states whose Bell violations can be arbitrarily enlarged by increasing the dimensions. Similarly, for superactivation of monogamy relations, by increasing the number $m$ of the copies, the monogamy relations can be always superactivated for nonadditive quantum correlation, e.g, under the regularized quantum correlation measure. From Fig. 1, one sees that the residual correlation can be made arbitrarily large by increasing the number of copies.

\section{Super activation of monogamy relations for multipartite systems}

Theorem 1 can be generalized to the multipartite systems. For any $n$-partite state $\rho_{A_1A_2\cdots A_n}$, one can view $A_1$, $A_2$, and $A_3\cdots A_n$ as $A$, $B$, and $C$, respectively, in Theorem 1. Then by partitioning the last system $C$ into two subsystems $A_3$ and $A_4\cdots A_n$, one can apply Theorem1 repeatedly. We have the following result.

{\bf[Theorem 3]}. Let $Q$ be a continuous measure of quantum correlation. For any state $\rho_{A_1A_2\cdots A_n}$, there exist real numbers $\alpha$ and $\beta$ such that

(1) if $x\geq \alpha$, then $Q^x$ is monogamous,
\begin{eqnarray}\label{th21}
Q^{x}_{A_1|A_2\cdots A_n}\geq \sum_{j=2}^nQ^{x}_{A_1A_j};
\end{eqnarray}

(2) if $0\leq y\leq \beta$, then $Q^y$  is polygamous,
\begin{eqnarray}\label{th22}
Q^{y}_{A_1|A_2\cdots A_n}\leq \sum_{j=2}^nQ^{y}_{A_1A_j}.
\end{eqnarray}

Similarly concerning Theorem 2, we have for multipartite system:

{\bf[Theorem 4]}. Let $Q$ be a continuous measure of quantum correlation.
For any state $\rho_{A_1A_2\cdots A_n}$ that $Q^x$ does not satisfy the monogamy relation (\ref{th21}) for $\beta\leq x\leq \alpha$, there always exists a positive integer $m$ such that
\begin{eqnarray}\label{th4n}
Q(\rho^{\otimes m}_{A_1|A_2\cdots A_n})\geq \sum_{j=2}^n Q(\rho^{\otimes m}_{A_1A_j}).
\end{eqnarray}

Let us consider the entanglement measure negativity $N_c$, and a class of $n$-qudit quantum states $|W^d_n\rangle_{A_1\cdots A_n}=\sum_{i=1}^{d-1}(a_{1i}|i0\cdots0\rangle+\cdots+a_{ni}|00\cdots i\rangle)$, with $\sum_{s=1}^{n}\sum_{i=1}^{d-1}|a_{si}|^2=1$.
First, we prove the negativity $N_c$ of $|W_n^d\rangle_{A_1A_2\cdots A_n}$ does not satisfy the usual monogamy inequality (\ref{monogamy}). Then, we find the positive integer $m$ such that the monogamy relation for $|W_n^d\rangle^{\otimes m}_{A_1A_2\cdots A_n}$ is established.

The reduced density matrix $\rho_{A_1}$ of $|W_n^d\rangle_{A_1A_2\cdots A_n}$ with respect to the subsystem $A_1$ is given by
\begin{eqnarray}\label{npfth11}
\rho_{A_1}&&=\mathrm{Tr}_{A_2\cdots A_n}(|W_n^d\rangle_{A_1A_2\cdots A_n}\langle W_n^d|)\nonumber\\
&&=\sum_{i,j=1}^{d-1}a_{1i}a_{1j^\ast}|i\rangle_{A_1}\langle j|+\Omega_1|0\rangle_{A_1}\langle 0|,
\end{eqnarray}
where $\Omega_1=\sum_{s=2}^n\sum_{i=1}^{d-1}|a_{si}|^2=1-\sum_{j=1}^{d-1}|a_{1j}|^2$.
From the definition of the negativity and Eq. (\ref{npfth11}), we have
\begin{eqnarray}\label{npfth12}
N_c(|W_n^d\rangle_{A_1|A_2\cdots A_n})&&=(\mathrm{Tr}\sqrt{\rho_{A_1}})^2-1\nonumber\\
&&=2\sqrt{(1-\Omega_1)\Omega_1}.
\end{eqnarray}

The two-qudit reduced density matrix $\rho_{A_1A_2}$ of $|W_n^d\rangle_{A_1A_2\cdots A_n}$,
similar for $\rho_{A_1 A_s})$, $s=2,\cdots, n$, is given by
\begin{eqnarray}\label{npfth13}
\rho_{A_1A_2}&&=\mathrm{Tr}_{A_3\cdots A_n}(|W_n^d\rangle_{A_1A_2\cdots A_n}\langle W_n^d|)\nonumber\\
&&=\sum_{i,j=1}^{d-1}(a_{1i}a_{1j^\ast}|i0\rangle_{A_1A_2}\langle j0|+a_{1i}a_{2j^\ast}|i0\rangle_{A_1A_2}\langle 0j|\nonumber\\
&&+a_{2i}a_{1j^\ast}|0i\rangle_{A_1A_2}\langle j0|+a_{2i}a_{2j^\ast}|0i\rangle_{A_1A_2}\langle 0j|)\nonumber\\
&&+\Omega_2|00\rangle_{A_1A_2}\langle 00|,
\end{eqnarray}
where $\Omega_2=1-\sum_{j=1}^{d-1}(|a_{1j}|^2+|a_{2j}|^2)$. Note that, by using the following two un-normalized states
\begin{eqnarray}\label{npfth14}
&&|\hat{x}\rangle_{A_1A_2}=\sum_{i=1}^{d-1}(a_{1i}|i0\rangle_{A_1A_2}+a_{2i}|0i\rangle_{A_1A_2}),\nonumber\\
&&|\hat{y}\rangle_{A_1A_2}=\sqrt{\Omega_2}|00\rangle_{A_1A_2},
\end{eqnarray}
$\rho_{A_1A_2}$ in Eq. (\ref{npfth13}) can be represented as
\begin{eqnarray}\label{npfth15}
\rho_{A_1A_2}=|\hat{x}\rangle_{A_1A_2}\langle\hat{x}|+|\hat{y}\rangle_{A_1A_2}\langle\hat{y}|.
\end{eqnarray}

For any pure state decomposition
\begin{eqnarray}\label{npfth16}
\rho_{A_1A_2}=\sum_{h}|\hat{\phi}_h\rangle_{A_1A_2}\langle\hat{\phi}_h|,
\end{eqnarray}
where $|\hat{\phi}_h\rangle_{A_1A_2}$ is an un-normalized state in two-qudit subsystem $A_1A_2$, there exists a unitary matrix $u$ with entries $u_{hl}$ such that
\begin{eqnarray}\label{npfth17}
|\hat{\phi}_h\rangle_{A_1A_2}=u_{h1}|\hat{x}\rangle_{A_1A_2}+u_{h2}|\hat{y}\rangle_{A_1A_2},
\end{eqnarray}
for each $h$. For the normalized state $|\phi_h\rangle_{A_1A_2}=|\hat{\phi}_h\rangle_{A_1A_2}/\sqrt{p_h}$ with $p_h=|\langle \hat{\phi}_h|\hat{\phi}_h\rangle|$, the pure state negativity is given by
\begin{eqnarray}\label{npfth18}
N_c(|\phi_h\rangle_{A_1A_2})&&=\frac{2}{p_h}|u_{h2}|^2\sqrt{(1-\Omega_1)(\Omega_1-\Omega_2)}\nonumber\\
&&=\frac{2}{p_h}|u_{h2}|^2\sqrt{(1-\Omega_1)\sum_{i=1}^{d-1}|a_{2i}|^2}
\end{eqnarray}
for each $h$.

From  Eq. (\ref{nc}) and Eq. (\ref{npfth18}), we have
\begin{eqnarray}\label{npfth19}
N_c(\rho_{A_1A_2})&&=\min_{\{p_h,|\phi_h\rangle\}}\sum_hp_hN_c(|\phi_h\rangle)_{A_1A_2}\nonumber\\
&&=\min_{\{p_h,|\phi_h\rangle\}}\sum_h2|u_{h2}|^2\sqrt{(1-\Omega_1)\sum_{i=1}^{d-1}|a_{2i}|^2}\nonumber\\
&&=2\sqrt{(1-\Omega_1)\sum_{i=1}^{d-1}|a_{2i}|^2},
\end{eqnarray}
where the last equality is due to the choice of $u_{h2}$ from the unitary matrix $u$.
Here we note that the minimum average of the CREN in Eq. (\ref{npfth19}) does not depend on the choice of the pure state decomposition of $\rho_{A_1A_2}$, which simplifies the minimization problem.

By using an analogous method, we have
\begin{eqnarray}\label{npfth110}
N_c(\rho_{A_1A_s})=2\sqrt{(1-\Omega_1)\sum_{i=1}^{d-1}|a_{si}|^2},
\end{eqnarray}
for each $s=2,\cdots,n$.

From Eq. (\ref{npfth12}) and Eq. (\ref{npfth110}), one can easily obtain that $N_c(|W_n^d\rangle_{A_1|A_2\cdots A_n})\leq \sum_{s=2}^nN_c (\rho_{A_1A_s})$. Therefore, $N_c$ is not monogamous for the state $|W_n^d\rangle_{A_1A_2\cdots A_n}$.

Now let us consider $m$ copies of the state $|W_n^d\rangle_{A_1A_2\cdots A_n}$, $|W_n^d\rangle_{A_1A_2\cdots A_n}^{\otimes m}$.
We get $N_c(|W_n^d\rangle_{A_1|A_2\cdots A_n}^{\otimes m})=\frac{1}{2}\big[(1+4\sqrt{(1-\Omega_1)\Omega_1})^m-1\big]$, and $N_c(\rho_{A_1A_s}^{\otimes m})=\frac{1}{2}\big[(1+4\sqrt{(1-\Omega_1)\sum_{i=1}^{d-1}|a_{si}|^2})^m-1\big]$ for $s=2,\cdots,n$.
Since $1+4\sqrt{(1-\Omega_1)\Omega_1}>1+4\sqrt{(1-\Omega_1)\sum_{i=1}^{d-1}|a_{si}|^2}>1$, $s=2,\cdots,n$, it is always possible to choose some positive integer $m$ such that the monogamy relation holds.
For example, let $d=2$, $n=5$, $a_{11}=\cdots=a_{51}=\frac{1}{\sqrt{5}}$. Then $N_c(|W_5^2\rangle_{A_1|A_2\cdots A_5}^{\otimes m})=\frac{1}{2}[(\frac{13}{5})^m-1]$, and $N_c(\rho_{A_1A_s}^{\otimes m})=\frac{1}{2}[(\frac{9}{5})^m-1]$ for $s=2,\cdots,5$.
It is easy to prove that the monogamy relation is satisfied for $m\geq 4$, see Fig. 2.
\begin{figure}
  \centering
  \includegraphics[width=9cm]{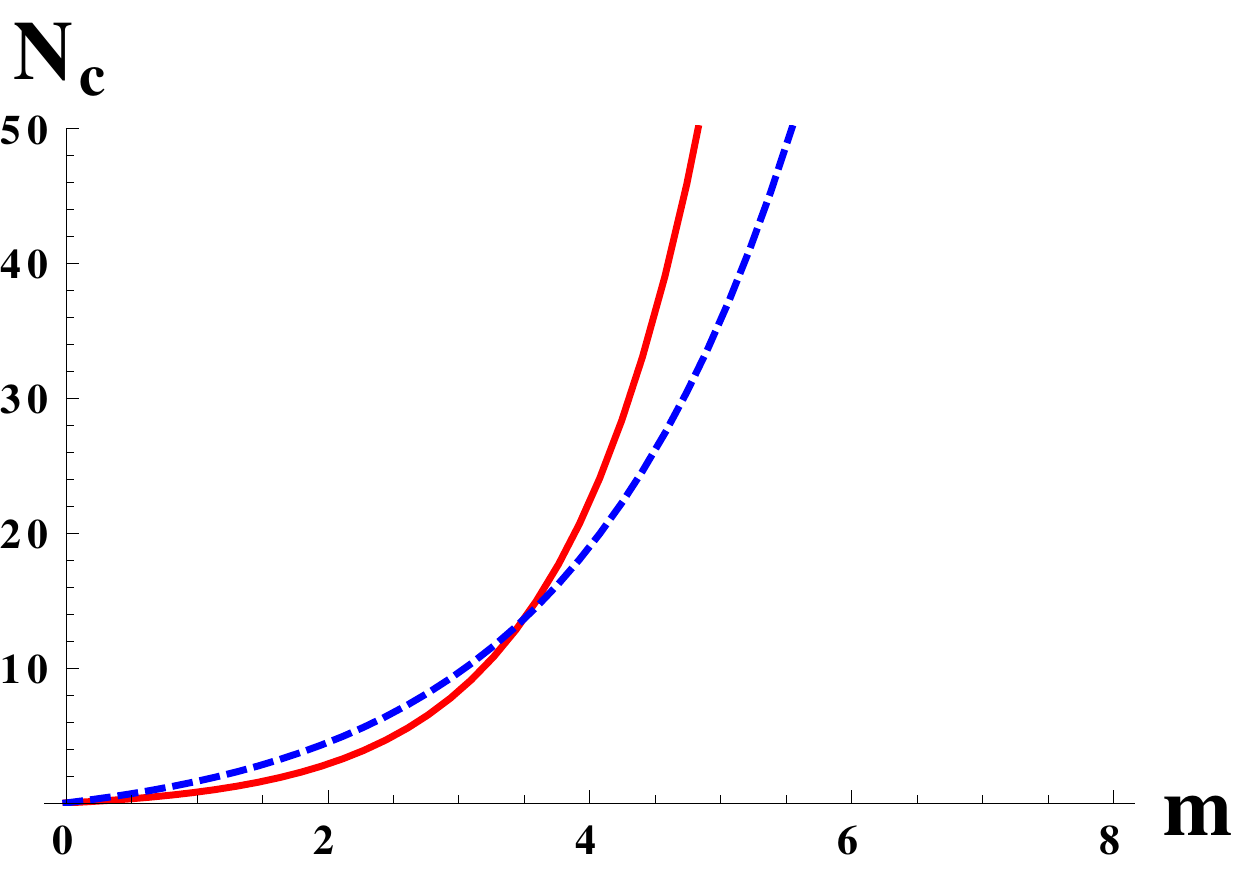}\\
  \caption{Solid (red) line is the negativity $N_c(|W_5^2\rangle_{A_1|A_2\cdots A_5})$; dashed (blue) line is the summation of negativity of $\rho_{A_1A_s}$, $s=2,\cdots,5$.}\label{1}
\end{figure}

To illustrate the relations among the superactivation of monogamy property, the number $m$ of the copies and the number $n$ of the subsystems, let us consider the case $a_{ij}=\frac{1}{\sqrt{nd}}$, $i=1,2,...,n$, $j=1,2,...,d$. We have $N_c(|W_{n}^{d}\rangle_{A_1|A_2\cdots A_{n}}^{\otimes m})=\frac{1}{2}[(1+\frac{4\sqrt{n-1}}{n})^m-1]$, and $N_c(\rho_{A_1A_s}^{\otimes m})=\frac{1}{2}[(1+\frac{4}{n})^m-1]$ for each $s=2,\cdots,n$.
Therefore, $N_c(|W_{n}^{d}\rangle_{A_1|A_2\cdots A_{n}}^{\otimes m})-
\sum_{s=2}^n N_c(\rho_{A_1A_s}^{\otimes m})=
\frac{1}{2}[(1+\frac{4\sqrt{n-1}}{n})^m-(n-1)(1+\frac{4}{n})^m+n-2]\equiv f(n,m)$.
Figure 3 shows that to keep $f(n,m)\geq 0$, with the increase of the number $n$ of subsystems, the number $m$ of the copies should also increase. When $n$ tends to infinity, $f(n,m)$ is 0 for any $m$. If $m$ is sufficiently large, the monogamy relation is always satisfied.
Here, from the conclusion of Theorem 2, the regularized quantum correlation measure always satisfies the monogamy relation, $Q^\infty(\rho^{\otimes m}_{A_1|A_2\cdots A_n})\geq \sum_{j=2}Q^\infty(\rho^{\otimes m}_{A_1A_j})$.

\begin{figure}
  \centering
  \includegraphics[width=9cm]{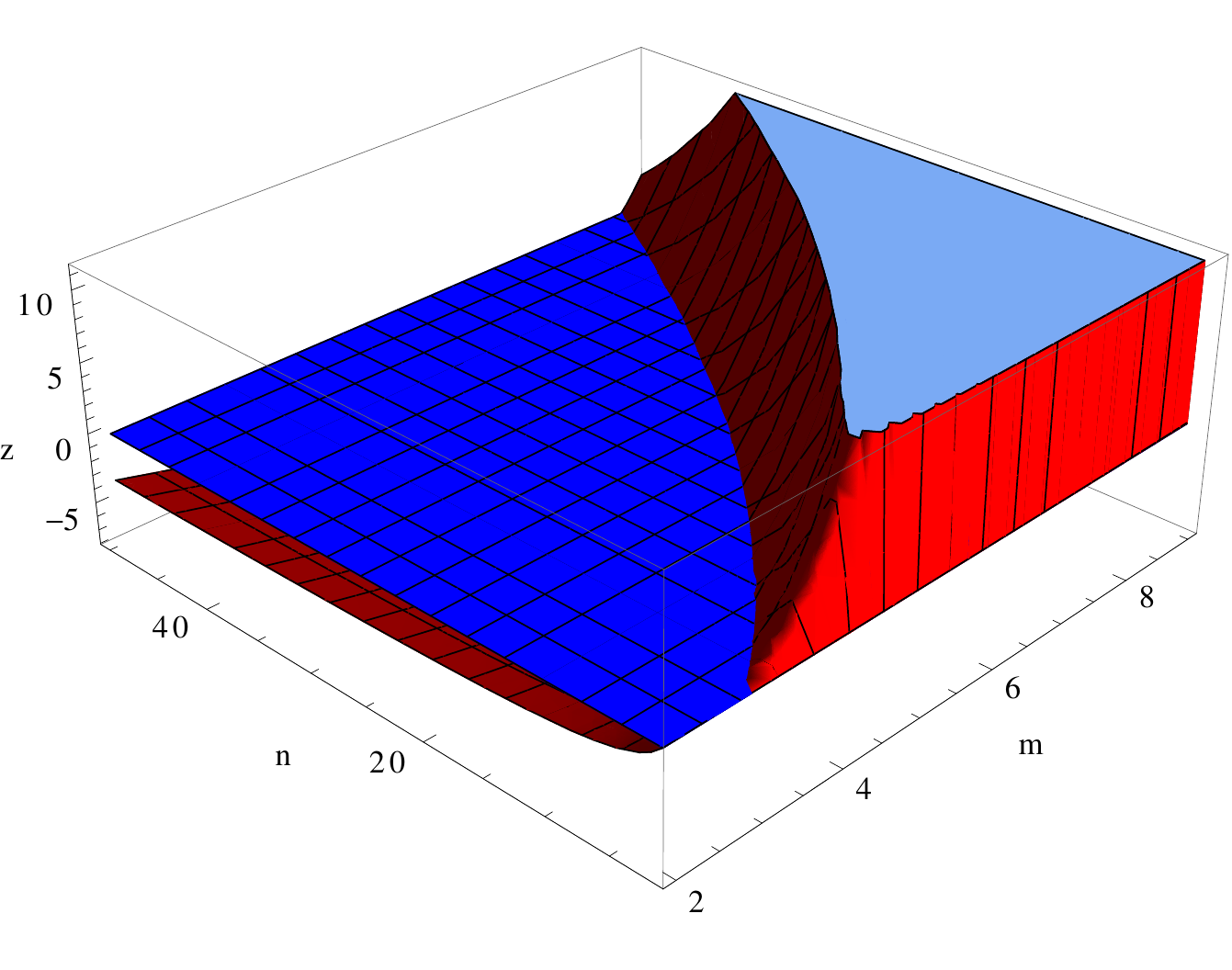}\\
  \caption{$m$ is the number of the copies; $n$ is the number of the subsystems; $z$ represents the value of the function $f(n,m)$ (red surface). The blue surface is the zero plane of $z$.}\label{1}
\end{figure}

\section{Tighter monogamy relation of multipartite systems}

In the following, we show that these monogamy inequalities satisfied by the quantum correlation measures can be further refined and become tighter.

{\bf[Theorem 5]}.
Let $Q$ be a continuous measure of quantum correlation. For any tripartite state $\rho_{ABC}$, if $Q_{AB}\geq Q_{AC}$, then for any real number $s\geq 1$, we have
\begin{eqnarray}\label{th31}
Q^t_{A|BC}\geq Q^t_{AB}+(2^{s}-1)Q^t_{AC},
\end{eqnarray}
where $t=s\alpha$, where $\alpha$ is given in Theorem 1.

{[\sf Proof]}. By using the inequality $(1+t)^x\geq 1+(2^{x}-1)t^x$, $0\leq t \leq 1$, $x\in [1,\infty)$, we have
\begin{eqnarray*}
  Q^{xs}_{A|BC}
  &&\geq (Q^x_{AB}+Q^x_{AC})^s\\
  &&=Q^{xs}_{AB}\left(1+\frac{Q^x_{AC}}{Q^x_{AB}}\right)^s \\
  && \geq Q^{xs}_{AB}\left[1+(2^s-1)\left(\frac{Q^x_{AC}}{Q^x_{AB}}\right)^s\right]\\
  &&=Q^{xs}_{AB}+(2^s-1)Q^{xs}_{AC}.
\end{eqnarray*}
As the subsystems $A$ and $B$ are equivalent in this case, we have assumed that $Q_{AB}\geq Q_{AC}$ without loss of generality. Moreover, if $Q_{AB}=0$, we have $Q_{AB}=Q_{AC}=0$: The lower bound becomes trivially zero. \hfill \rule{1ex}{1ex}

Since for $s\geq 1$, $2^s-1\geq1$, (\ref{th31}) in Theorem 5 gives a tighter monogamy relation, with larger lower bounds, than (\ref{th11}) in Theorem 1.
For the multipartite state, we can also have a tighter monogamy relation.

{[\bf Theorem 6]}. For multipartite state $\rho_{A_1A_2\cdots A_n}\in $ and a continuous measure $Q$ of quantum correlation, if
$Q_{A_1A_i}\geq Q_{A_1|A_{i+1}\cdots A_n}$ for $i=2, \cdots, m$, and
$Q_{A_1A_j}\leq Q_{A_1|A_{j+1}\cdots A_n}$ for $j=m+1,\cdots,n-1$,
$\forall$ $2\leq m\leq n-2$, $n\geq 4$, we have
\begin{eqnarray}\label{th4}
&&Q^t_{A_1|A_2\cdots A_n}\geq \nonumber \\
&&Q^t_{A_1A_2}+(2^s-1) Q^t_{A_1A_3}+\cdots+(2^s-1)^{m-2}Q^t_{A_1A_m}\nonumber\\
&&+(2^s-1)^{m}(Q^t_{A_1A_{m+1}}+\cdots+Q^t_{A_1A_{n-1}}) \nonumber\\
&&+(2^s-1)^{m-1}Q^t_{AB_n}
\end{eqnarray}
for $t=s\alpha$, $s\geq1$, with $\alpha$ given in Theorem 1.

{\sf [Proof].} From the inequality (\ref{th31}), we have
\small
\begin{eqnarray}\label{pf41}
&&Q^t_{A_1|A_2\cdots A_n}\nonumber\\
&&\geq  Q^t_{A_1A_2}+(2^s-1)Q^t_{A_1|A_3\cdots A_n}\nonumber\\
&& \geq \cdots\nonumber\\
&&\geq Q^t_{A_1A_2}+(2^s-1)Q^t_{A_1A_3}+\cdots+(2^s-1)^{m-2}Q^t_{A_1A_m}\nonumber\\
&&~~~~+(2^s-1)^{m-1} Q^t_{A_1|A_{m+1}\cdots A_n}.
\end{eqnarray}
\normalsize
Similarly, as $Q_{A_1A_j}\leq Q_{A_1|A_{j+1}\cdots A_n}$ for $j=m+1,\cdots,n-1$, we get
\begin{eqnarray}\label{pf42}
&&Q^t_{A_1|A_{m+1}\cdots A_n} \nonumber\\
&&\geq (2^s-1)Q^t_{A_1A_{m+1}}+Q^t_{A_1A_{m+2}\cdots A_n}\nonumber\\
&&\geq (2^s-1)(Q^t_{A_1A_{m+1}}+\cdots+Q^t_{A_1A_{n-1}})\nonumber\\
&&~~~~+Q^t_{A_1A_n}.
\end{eqnarray}
Combining (\ref{pf41}) and (\ref{pf42}), we have Theorem 6. \hfill \rule{1ex}{1ex}

\section{conclusion}
Entanglement monogamy and polygamy are fundamental properties of quantum multipartite states. We have studied the monogamy and polygamy relations related to general measures of quantum correlation. We have shown that there always exist two real numbers $\alpha$ and $\beta$ for any nonadditive quantum correlation measures and any given quantum state: $Q^x$ is monogamous if $x\geq \alpha$ and polygamous if $0\leq x\leq \beta$.

For $\beta <x<\alpha$, depending on the detailed quantum correlation measures, quantum states may satisfy neither the monogamy nor the polygamy relations.
However, similar to the nonlocality revealed by the violation of Bell inequalities,
where for some bipartite states that admit local hidden variable models, their nonlocality can superactivated by $m$ copies of the states \cite{supera}, we have shown that for a state which does not admit monogamous relations under $Q^x$, $\beta <x<\alpha$, the $m$ copies of the state could satisfy monogamous relations.
We have used the negativity as the quantum correlation measure to illustrate such superactivation of monogamy properties. Concerning infinitely many copies we have the regularized quantum correlation measures satisfying the monogamy relations. A tighter monogamy relation has also been given to any quantum correlation measures.

\bigskip
\noindent{\bf Acknowledgments}\, \, This work is supported by the Natural Science Foundation of China(NSFC) under Grants No. 11847209 and No. 11675113; Key Project of Beijing Municipal Commission of Education under No. KZ201810028042.


\begin{thebibliography}{99}
\bibitem{VJW} V. Coffman, J. Kundu, and W. K. Wootters, Phys. Rev. A 61, 052306 (2000).
\bibitem{FMA} F. Mintert, M. Ku\'s, and A. Buchleitner, Phys. Rev. Lett. 92, 167902 (2004).
\bibitem{KSS} K. Chen, S. Albeverio, and S. M. Fei, Phys. Rev. Lett. 95, 040504 (2005).
\bibitem{HPB} H. P. Breuer, J. Phys. A: Math. Gen. 39, 11847 (2006).
\bibitem{HPBB}H. P. Breuer, Phys. Rev. Lett. 97, 080501 (2006).
\bibitem{JIV} J. I. de Vicente, Phys. Rev. A 75, 052320 (2007).
\bibitem{CYS} C. J. Zhang, Y. S. Zhang, S. Zhang, and G. C. Guo, Phys. Rev. A 76, 012334 (2007).
\bibitem{MP} M. Pawlowski, Phys. Rev. A 82, 032313 (2010).	

\bibitem{jmr} J. M. Renes and M. Grassl, Phys. Rev. A 74, 022317 (2006).
 \bibitem{ml} L. Masanes, Phys. Rev. Lett. 102, 140501 (2009).
\bibitem{max} X. S. Ma, B. Dakic, W. Naylor, A. Zeilinger, and P. Walther, Nat. Phys. 7, 399 (2011).
\bibitem{kjm} K. Meichanetzidis, J. Eisert, M. Cirio, V. Lahtinen, and J. K. Pachos, Phys. Rev. Lett. 116, 130501 (2016).
\bibitem{aad} A. Almheiri, D. Marolf, J. Polchinski, and J. Sully, J. High Energy Phys. 02, 062 (2013)
\bibitem{jbe} J. Bae and A. Ac$\acute{i}$n, Phys. Rev. Lett. 97, 030402 (2006).
\bibitem{vss} V. Scarani, S. Iblisdir, N. Gisin, and A. Ac$\acute{i}$n, Rev. Mod. Phys. 77, 1225 (2005).



\bibitem{byk1}Y. K. Bai, M. Y. Ye, and Z. D. Wang, Phys. Rev. A 80, 044301(2009).
\bibitem{ZXN} X. N. Zhu and S. M. Fei, Phys. Rev. A 90, 024304 (2014).
\bibitem{byk2} Y. K. Bai, Y. F. Xu, and Z. D. Wang, Phys. Rev. Lett. 113, 100503 (2014).
\bibitem{JZX} Z. X. Jin and S. M. Fei, Quantum Inf Process  16:77 (2017).
\bibitem{jll}  Z. X. Jin, J. Li, T. Li, S. M. Fei, Phys. Rev. A 97, 032336 (2018).
\bibitem{j012334} J. S. Kim, Phys. Rev. A 97, 012334 (2018).
\bibitem{042332} J. S. Kim, Phys. Rev. A 97, 042332 (2018).
\bibitem{gy1} G. Gour, Y. Guo, Quantum 2, 81 (2018).

\bibitem{fgj}F. Buscemi, G. Gour, and J. S. Kim, Phys. Rev. A 80, 012324 (2009).
\bibitem{sa} S. Bagchi and A. K. Pati, Phys. Rev. A 91, 042323 (2015).
\bibitem{jzx1} Z. X. Jin and S. M. Fei, Quantum Inf Process  17:2 (2018).
\bibitem{jzx2}Z. X. Jin, S. M. Fei and X. Q. Li-Jost. Quantum Inf Process  17:213 (2018).
\bibitem{gy2} Y. Guo, Quantum Inf Process, 17:222 (2018).
\bibitem{fh} Y. C. Ou, H. Fan, Phys. Rev. A 75, 062308 (2007).
\bibitem{bcs} J. S. Kim, A. Das, and B. C. Sanders, Phys. Rev. A 79, 012329 (2009).
\bibitem{ycs} C. S. Yu and H. S. Song, Phys. Rev. A 77, 032329 (2008).
\bibitem{062302}J. S. Kim, Phys. Rev. A 85, 062302 (2012).
\bibitem{gpw}  B. Groisman,  S. Popescu,  A. Winter, Phys. Rev. A 72, 032317 (2005).
\bibitem{GSB} G. Goura, S. Bandyopadhyayb, and B. C. Sandersc, J. Math. Phys. 48, 012108 (2007).

\bibitem{fsm} Z. G. Li, S. M. Fei, S. Albeverio, W. M. Liu, Phys. Rev. A 80, 034301 (2009).
\bibitem{supera} C. Palazuelos, Phys. Rev. Lett., 109, 190401 (2012).
\bibitem{oyc} Y. C. Ou, Phys. Rev. A 75, 034305 (2007).
\bibitem{ak} A. Kumar, Phys. Lett. A 380, 3044–3050 (2016).

\bibitem{tj} T. J. Osborne and F. Verstraete, Phys. Rev. Lett., 96, 220503 (2006).
\bibitem{wj} X. J. Ren and W. Jiang, Phys. Rev. A, 81, 024305 (2010).
\bibitem{ksb} S. Karmakar, A. Sen, A. Bhar, and D. Sarkar, Phys. Rev. A 93, 012327 (2016).
\bibitem{GRF} G. Vidal and R. F. Werner, Phys. Rev. A. 65, 032314 (2002).

\bibitem{ly} Y. Luo and Y. Li, Ann. Phys, 362:511-520 (2015).
\bibitem{gdl}Y. Guo, S. Du, X. Li, S. J. Wu, Phys. A Math. Theor. 48, 245301 (2015).
\bibitem{gjl}Y. Guo,  Y. Jia,  X. Li, Quantum Inf. Process. 14, 3553 (2015).
\end{thebibliography}
\end{document}